\documentclass[letterpaper,12pt]{article}   

\usepackage{graphics, epsfig}

\begin{document}

\title{\bf Reciprocity constraints on the matrix of reflection from optically anisotropic surfaces}
\vspace{40mm}

\author{\bf Rajendra~Bhandari}
\date{}
\maketitle
\vspace{40mm}
\begin{center}
\begin{tabular}{ll}
            & Raman Research Institute, \\
            & Bangalore 560 080, India. \\
            & email: bhandari@rri.res.in\\
\end{tabular}
\end{center}
\vspace{35mm}
\vspace{25mm}

OCIS Codes 260.5430, 240.2130, 230.5440, 230.1360

\newpage

\section{Abstract}

We derive certain constraints on the reflection matrix for reflection from a plane, nonmagnetic, 
optically anisotropic surface using a reciprocity theorem stated long ago by van de Hulst in the context of 
scattering of polarized light. The constraints are valid for absorbing and chiral media and can  be used as 
tools to check the consistency of  
derived expressions for such matrices in terms of the intrinsic parameters of the reflecting 
medium as illustrated by several examples. 
 
\section{Introduction}

Except in the simplest situations where linearly polarized light is incident 
on an optically isotropic surface or a specially oriented anisotropic surface,  
the polarization state of a light wave in general changes when 
it is reflected off a surface. Changes in polarization states imply changes of 
phase, as we know from the work of Pancharatnam.  A correct analysis of optical devices  
involving polarization, with or without interference, requires a precise 
and reliable method for handling polarization transformations resulting from 
reflections. These are described by means of a 2 x 2 complex reflection matrix. 
The specification of this matrix requires the choice of a set of basis states 
for the incident as well as the reflected waves alongwith their phases. In this communication we 
first describe the most natural convention for the basis states and a consistent 
one for the description of reciprocity. We then derive certain constraints that 
reflection amplitudes must satisfy on account of the principle of reciprocity 
and illustrate their use with several examples. The principle of reciprocity in polarization 
optics has been widely discussed in literature and is different from time-reversal 
invariance in that (i) it applies to systems with absorption and (ii) it deals with one 
incident wave and one scattered wave at a time. For a recent review we refer the reader 
to Potton \cite{potton}. For the purpose of this paper the most appropriate formulation of 
reciprocity is the one given by van de Hulst 
\cite{vandehulst} wherein he choses to rotate the scatterer instead of reversing of the wave. 

\section{The phase convention}

It was pointed out  recently \cite{transpsymm,transpapp} that when a plane wave of light 
changes its direction of propagation from a wave vector $\vec k$ to a 
wave vector $\vec k'$ due to refraction, reflection or scattering, the definition 
of the matrix that relates the incident polarization state to the final polarization state, i.e. 
the Jones matrix,  requires the choice of a set of basis states and their phases for each of the 
two directions of propagation. The  choice that is most often used both in polarization 
optics as well as in scattering theory is the following, also illustrated in Figs. 1 and 2. 

For the wave vector $\vec k$, a 
set of orthogonal, linearly polarized states along $\hat x$ and $\hat y$, 
called the p and the s states, are chosen as the base states 1 and 2 respectively 
where $\hat x$ is in the plane and $\hat y$ is 
perpendicular to the plane of reflection or scattering. The  phases of the basis states are 
chosen such that in the basis state 1, $~E_x=E{\rm exp}(i\omega t),~ E_y=0 ~$ and 
in the basis state 2 $~ E_x=0, ~E_y=E{\rm exp}(i\omega t)$,
where $E_x$ and $E_y$ are the $x$ and $y$ components of the electric field in the wave. 
With this convention, the vector $(1/\sqrt 2)$col.[1,1] 
represents a linearly polarized state along a direction lying in the $(\hat x ,\hat y)$ plane, 
making an angle $45^\circ$ with $\hat x$ and the vectors $(1/\sqrt 2)$col.[1,$\pm i$] 
represent the right and left circularly polarized states. For the wave vector $\vec k'$, 
the convention most often used in scattering theory as well as in polarization optics 
is the following: Rotate the basis states about an axis perpendicular to the plane of 
reflection, i.e. along $\hat y$, through an angle such that $\vec k$ goes to $\vec k'$. 
Let $\hat x$ and $\hat y$ go to $\hat x'$ and $\hat y'$ under this rotation. The polarization 
basis states for $\vec k'$ are chosen to be linearly polarized states along $\hat x'$ and $\hat y'$
with their relative phases chosen such that in the basis state 1, $~E_x'=E{\rm exp}(i\omega t),~ E_y'=0 ~$ and 
in the basis state 2 $~ E_x'=0, ~E_y'=E{\rm exp}(i\omega t)$. 
Note that since the rotation is about $\hat y$, $\hat y'$ = $\hat y$.
This choice leads to a precise phase convention for back-scattering 
or for reflection at normal incidence at a surface, where $\vec{k'}$=$-\vec k$. 
For this case one gets $\hat x'$ = $-\hat x$ and $\hat y'$ = $\hat y$. 
We shall call the above convention the ``travelling-frame convention". 
In literature the above described choice of phase convention is referred to as ``choice of the 
coordinate system".

\section{The reciprocity constraints}

Scattering of a polarized plane wave with wave-vector $\vec k$ to a wave with wave-vector $\vec k'$ from a scatterer 
is described by a  2 x 2 complex scattering matrix $A$ whose  matrix element $A_{ij}$ represents the 
complex amplitude for an incident wave with unit amplitude in polarization state $i$ to 
scatter into the polarization state $j$.
If $A(\vec k,\vec k')$ represents the  matrix for scattering  
from $\vec k$ to $\vec k'$  and $A(-\vec k',-\vec k)$ 
the matrix for the reverse scattering, the correct statement 
of the principle of reciprocity with the above phase convention has been given by Sekera \cite{sekera} 
as

\begin{eqnarray}
 & & A(-\vec k',-\vec k)= \bar A(\vec k,\vec k') \label{eq:r1}
\end{eqnarray}

\noindent where the matrix $\bar A$ is the ``n-transpose" of $A$, defined  \cite{transpsymm} as,

\begin{eqnarray}
\bar A_{ij} = (-1)^{i+j}A_{ji}& & \label{eq:r2}
\end{eqnarray}

\noindent For a 2 x 2 matrix, $\bar A$ is the transpose of $A$ with a change of  sign of the off-diagonal elements.

In order to derive the constraints on the reflection matrix resulting from 
reciprocity, a somewhat different formulation of the principle of reciprocity, first made by 
van de Hulst \cite{vandehulst} in the context of scattering problems, is more useful \cite{footnote1}. van de Hulst 
choses to rotate the scatterer 
rather than reverse the direction of propagation. Let the polarization basis states for  
$\vec k$ and $\vec k'$ be defined using the travelling-frame convention, let $\vec k$ and $\vec k'$ 
be along $\hat z$ and $\hat z'$ respectively and let the scattering be in the ($\hat x$,$\hat z$) 
plane as shown in Fig. 1 where $\alpha$ is the scattering angle.
van de Hulst's theorem states: If the scatterer is rotated through $180^\circ$ 
about an axis defined by the line bisecting the angle between the vectors $\vec k'$ and 
$-\vec k$ , called the bisectrix, the matrix $A(\vec k,\vec k')$ goes to the 
matrix $\bar A(\vec k,\vec k')$ where $\bar A$ is the 
n-transpose of $A$ defined by Eq.(\ref{eq:r2}).

The theorem, translated to the problem of reflection from a plane surface in optics, can be phrased 
as follows: If the reflecting medium, assumed to be reciprocal, 
is rotated about the normal $\hat n$ to the surface SS
through $180^\circ$ (Fig.2), the reflection matrix $Z$ goes to $Z'$ where

\begin{eqnarray}
Z'~=~ \bar Z , & & \label{eq:r3}
\end{eqnarray}

\noindent where $\bar Z$ is the 
n-transpose of $Z$ defined by Eq.(\ref{eq:r2}).

The theorem is true in the presence of absorption and dichroism and has the straightforward 
consequence that if the reflecting medium, assumed to be reciprocal, is 
invariant under a rotation through $\pi$ about $\hat n$, the reflection matrix $Z$, 
for any angle of incidence, must be antisymmetric. For such cases therefore,

\begin{eqnarray}
Z_{ij}=-Z_{ji}. & & \label{eq:r6}
\end{eqnarray}

\noindent Such cases include:

A. An optically isotropic medium, i.e. a medium with no birefringence or dichroism, 
linear or circular.

B. A medium with only optical activity and circular dichroism but no linear birefringence and 
no linear dichroism,

C. An absorbing uniaxial medium with or without optical activity,
with the optic axes for birefringence and dichroism coinciding  
and being perpendicular to the surface, 

D. An absorbing uniaxial medium with or without optical activity,
with the optic axes for birefringence and dichroism coinciding , 
and lying in  the plane of the surface, 

E. A nonabsorbing biaxial medium with one of the principal axes perpendicular 
to the surface.

In addition to the above cases, when light is incident normally, any reflecting medium is 
invariant under a $\pi$ rotation about the direction of the incident beam. 
At normal incidence therefore, the reflection matrix for any  reciprocal 
medium must be antisymmetric.

In cases A, B and C, when light is incident on the surface normally, there 
is an additional constraint when the reflecting surface is invariant under 
rotations about the direction of incidence, i.e about the normal to the surface.  
In the travelling frame convention this additional constraint can be expressed 
 as,

\begin{eqnarray}
R(\phi) Z  R(\phi) = Z, & & \label{eq:r7}
\end{eqnarray}

\noindent where 

\begin{eqnarray}
R(\phi) =
\left( 
  \begin{array}{lr}
\rm {cos} \phi & -\rm {sin}\phi \\
\rm {sin}\phi &\rm{cos}\phi\\
\end{array}
\right) & & \label{eq:r8}
\end{eqnarray}

\noindent is a rotation matrix that rotates an incident Jones vector by an arbitrary angle $\phi$.

Equation (\ref{eq:r7}) can be proved as follows:
Let the reflection matrix of the unrotated sample be $Z$ and let 
$\mid\psi_f>$ be the final state resulting from reflection of an incident state $\mid\psi_i>$ 
so that 

\begin{eqnarray}
\mid\psi_f>~=~ Z\mid\psi_i>& & \label{eq:r20}
\end{eqnarray}

The reflection matrix $Z^R$ of the rotated sample is obtained from the 
condition that when the  state $\mid\psi_i>$, rotated through an angle $\phi$, 
is incident on the rotated sample, the reflected state, in the travelling frame, 
must be the state $\mid\psi_f>$, rotated by an angle $-\phi$ i.e. 

\begin{eqnarray}
Z^R R(\phi)\mid\psi_i>~=~R(-\phi)\mid\psi_f> & & \label{eq:r21}
\end{eqnarray}

\noindent Eqns.(\ref{eq:r20}) and (\ref{eq:r21}) give 

\begin{eqnarray}
Z^R ~=~R(-\phi)Z R(-\phi)& & \label{eq:r22}
\end{eqnarray}

\noindent Since $\phi$ is arbitrary, requirement of invariance under rotation about 
the beam axis therefore gives Eq. (\ref{eq:r7}).

Let the matrix  that satisfies the  reciprocity constraint (\ref{eq:r6}), as well as the 
isotropy constraint (\ref{eq:r7}) be called $Z_0$. It can easily be shown that $Z_0$ must 
be of the form

\begin{eqnarray}
Z_0 = r
\left( 
  \begin{array}{lr}
1 & 0 \\
0 &-1\\
\end{array}
\right) .& & \label{eq:r9}
\end{eqnarray}

\noindent where $r$ is a complex number.  In other words whenever a reciprocal reflecting medium is 
invariant under an arbitrary rotation about the normal to the surface its reflection matrix 
for normal incidence is given by $Z_0$.  The result is known to be true for optically 
isotropic surfaces. The fact that it is true in the presence of optical activity and 
circular dichroism and that it follows from simple symmetry considerations came as a news 
to the author. We also note that since the matrix $Z_0$ 
is diagonal it is robust against a phase change between the two basis states.

\section{Dependence on the polarization basis}

We next discuss the  dependence of the theorem given  by Eq.(\ref{eq:r3}) on the basis 
states used to express the Jones vectors for the propagation directions $\vec k$ and $\vec k'$. 
Although while stating the theorem we assumed a basis of ``in-phase" linearly polarized states 
along $\hat x$ and $\hat y$, 
this is by no means the only possible choice for the theorem (\ref{eq:r3}) to be true. 
One can also choose as basis states, a pair of 
orthogonal elliptically polarized states with the principal axes of the polarization ellipses being
along $\hat x$ and $\hat y$; the states being phased such that at t=0, in the wave with wavevector $\vec k$,
the basis state 1 has   $E_y=0$ and the basis state 2 has $E_x=0$. Similarly, in the wave with  
wavevector $\vec {k'}$, at t=0,  $E_{y'}=0$ in basis state 1 and $E_x'=0$ in basis state 2. 
The theorem as stated above remains valid 
in this set of basis states. This can be proved easily.

The basis described above is obtained from the original linearly polarized basis by means 
of a unitary transformation U  that  takes the state $\mid {\hat x}>$ i.e. the state 
with coordinates $(90^\circ,0^\circ)$ on the Poincar\'{e} sphere, along a geodesic arc, to a state 
$\mid P>$ with coordinates $(90^\circ +\eta,0^\circ)$, where $ 90^\circ > \eta \geq -90^\circ$. 
Such a transformation is achieved by means of a linearly 
birefringent waveplate with retardation $\eta$,
with its fast axis at $45^\circ$ to the $\hat x$ axis. The matrix $U$ is 
therefore given by 

\begin{eqnarray}
 & & U = L_{45}(\eta)= R(45)L_0(\eta) R(-45). \label{eq:r10}
\end{eqnarray}

\noindent where $L_{\beta}(\eta)$ is the Jones matrix for a linear retarder with retardation $\eta$ and 
fast axis making an angle $\beta$ with $\hat x$. The fields in the basis states 1 and 2 
are given by, (1) $~E_x=E{\rm cos}(\eta/2){\rm exp}(i\omega t),~ E_y =-iE{\rm sin}(\eta/2){\rm exp}(i\omega t) ~$ and 
(2) $~ E_x=iE{\rm sin}(\eta/2){\rm exp}(i\omega t), ~E_y=E{\rm cos}(\eta/2){\rm exp}(i\omega t)$. 
The cases $\eta=0$ and $\eta=\pi /2$ give the fields in the linearly polarized  and 
the circularly polarized basis respectively.

The matrices $Z$ and $Z'$ when transformed to the new basis are given by 
$Q$ and $Q'$ where 

\begin{eqnarray}
 & & Q =  U Z U^\dag  ~~{\rm and}~~ Q' =  UZ' U^\dag  \label{eq:r11}
\end{eqnarray}

\noindent It can easily be shown by required matrix multiplication that 

\begin{eqnarray}
 & & Q' = \bar Q \label{eq:r12}
\end{eqnarray}

\noindent Equation (\ref{eq:r12}) states the reciprocity principle in the basis of the chosen 
elliptically polarized states. It needs to be mentioned however that in an elliptically 
polarized basis, the form of the matrix under conditions of normal incidence and spatial isotropy (\ref{eq:r7}) 
is not given by Eq.(\ref{eq:r9}). The latter requires $R(\phi)$ to be of the form (\ref{eq:r8}) 
which is true only in the linear basis.

\section{Applications}

The constraints derived above can be used as tools  
to check derived  expressions for the matrices of reflection from optically anisotropic surfaces 
in terms of the intrinsic parameters of the sample. Under appropriate conditions the derived 
expressions must satisfy these constraints. We cite below some examples from literature where derived 
expressions for reflection matrix elements indeed do so.

Sosnowski \cite{sosnowski} has derived the reflection matrix elements for reflection 
from the surface of a uniaxially anisotropic medium placed in an isotropic ambient medium 
for the case when the optic axis is parallel to the interface and is oriented at an 
angle $\alpha$ from the plane of incidence. These have been reproduced on p.355 of \cite{azzam}.
First consider the case of normal incidence i.e. $\phi_0=0$. We  derived the expressions for the off-diagonal 
elements $r_{ps}$ and $r_{sp}$ for this case using the formulae in Eqs. (4.244) - (4.246) of \cite{azzam}. It was found that 
they satisfy $r_{ps}=-r_{sp}$ as required by Eq.(\ref{eq:r6}) of this paper. Next consider the 
case of oblique incidence i.e. $\phi_0 \neq 0$. For this case we programmed the above chain of formulae  
on an Excel worksheet and computed $r_{ps}$ and $r_{sp}$ for several hundred randomly chosen sets 
of the parameters $N_0$, $N_{1o}$, $N_{1e}$, $\alpha$ and $\phi_0$ in their allowed ranges where $N_0$ is the 
refractive index of the isotropic ambient and $N_{1o}$, $N_{1e}$ are the two refractive 
indices of the anisotropic medium. In every case we obtained $r_{ps}=-r_{sp}$. To cite two specific
examples, for $n_0=1.2$, $n_{1o}=1.7$, $n_{1e}=1.3$, $\phi_0=30^\circ$ and $\alpha=60^\circ$, we 
obtained $r_{ps} = .0637$ and $r_{sp} = -.0637$ and for $n_0=1$, $n_{1o}=1.2$, $n_{1e}=1.5$, $\phi_0=75^\circ$ and $\alpha=10^\circ$, we 
obtained $r_{ps} = -.0274$ and $r_{sp} = .0274$.

Engelsen \cite{engelsen} has derived the expressions for the matrix of reflection from a uniaxially 
anisotropic film on an isotropic substrate in an isotropic ambient medium for the case where the optic axis 
of the uniaxial medium is perpendicular to its boundaries with the substrate and the ambient. 
These have been reproduced on pages 356-357 of \cite{azzam}. The matrix is diagonal in this case. 
We derived the expressions for the diagonal 
elements $r_{ss}$ and $r_{pp}$ for normal incidence using the formulae in Eqs. (4.249) - (4.257) 
of \cite{azzam}. It was found that 
they satisfy $r_{pp}=-r_{ss}$ as required by Eq.(\ref{eq:r9}) of this paper.

We next consider some examples from the literature on  
reflection from a reciprocal, isotropic chiral medium where  the constraints  derived in this paper  
yield useful insights.

Lekner \cite{lekner} has derived expressions for the reflection matrix for reflection from the 
boundary of an achiral and an isotropic chiral medium. For normal incidence these expressions 
yield $r_{pp}=r_{ss}$ and $r_{ps}=r_{sp}=0$. For oblique incidence the expressions satisfy 
$r_{ps}=r_{sp}$. These results differ from the results of this paper by a sign. The reason lies in the 
phase convention for the reflected $p$-wave used in \cite{lekner} (see section 1.3) which differs by $\pi$ from the one 
used in this paper (Fig. 2), resulting in a  change in the sign of the reflected $p$-wave amplitude. This changes the signs 
of $r_{sp}$ and $r_{pp}$. When this change of sign is accounted for, the results of \cite{lekner} 
agree with those of this paper.

Silverman \cite{silverman}  derived the reflection matrix for reflection at the surface 
of an isotropic, nonmagnetic chiral medium for two sets of constitutive relations that are (I) invariant 
and (II) noninvariant under a duality transformation of the electromagnetic fields. 
The symmetric constitutive relations (I) lead to null differential reflection at normal incidence for incident 
right and left-circularly polarized light.  The asymmetric constitutive relations (II)
on the other hand lead to non-zero differential reflection for right and left-circularly polarized light.
The author indicates a preference for (I) based on some difficulties 
with the results obtained from (II).
Using the constraint stated above, i.e. the reflection 
matrix for this case must be given by Eq. (\ref{eq:r9}), any theory that yields non-zero differential reflection 
at normal incidence for incident right and left-circularly polarized light can be ruled out. 
If we assume that the derivations in  \cite{silverman}, which do satisfy our constraints, are correct, 
it could be concluded  on grounds 
of symmetries alone that the asymmetric constitutive relations are incorrect. 

Georgieva \cite{georgieva}  reported a solution for the  amplitudes for reflection from the 
surface of a reciprocal optically active medium using a corrected Berreman's matrix, arguing that 
Berreman's matrix is incorrect since it yields unequal off-diagonal elements for the reflection matrix.
The considerations of this paper support this assertion. Interestingly however, while the off-diagonal 
elements in \cite{georgieva} are equal and opposite in sign as required by the above constraints, 
the diagonal elements do not satisfy these constraints when light is incident normally. 
Eqns.(27) and (30) in \cite{georgieva} yield, for normal incidence, $r_{ss}=r_{pp}$. The  constraint 
given by Eq.(\ref{eq:r9}) however requires $r_{ss}=-r_{pp}$. The negative sign is non-trival as it 
represents the difference between 
a plane glass plate and a halfwave plate. In this work since $r_{ss}=r_{pp}$ and $r_{ps}=-r_{sp}$, the 
disagreement cannot be explained by a phase convention. We believe therefore that there is a problem with 
the derivation in \cite{georgieva}. 

\section{Discussion}

Our reason for dwelling at length on the conventions regarding basis states is that the 
reflection matrix as well as the statement of the reciprocity principle depend on these 
conventions. While the travelling-wave convention is a fairly standard one, used 
by most textbooks on optics \cite{opticstexts,azzam} to relate the polarization states for $\vec k$ to those for 
$\vec k'$  for defining the reflection matrix, there are occasional exceptions. 
For example Lekner \cite{lekner} and Bassiri et al. \cite{bassiri} use a different convention 
which we shall call the ``fixed-frame convention". Consequently they
obtain  expressions for the Fresnel reflection amplitudes for reflection off a 
chiral surface that differ from those in \cite{opticstexts,azzam} even in the limit 
when the chiral parameter goes to zero. As mentioned before the amplitudes obtained 
with the two conventions are related by a change of sign of the amplitude of the 
reflected $p$-wave, hence of  $r_{sp}$ and $r_{pp}$.

In the analysis of propagation problems involving a series of oblique reflections 
terminating in a reflection at normal incidence so that the beam retraces its path, as for example 
in a Michelson interferometer, the problem of phase convention occurs twice, once 
while defining the reflection matrix and again while relating the forward and backward propagating 
waves. Since the first choice implies a choice for 
normal incidence, the natural thing to do is to make the second choice to be consistent 
with the first one. Unfortunately this has not always been the practice in literature. 
For example in Vansteenkiste et al.\cite{remotecontrol}   the travelling frame 
convention is used for the reflections at oblique incidence and a fixed-frame convention for 
relating the forward and backward propagating waves. As a consequence  
the  matrix for  reflection at normal incidence is defined differently 
from the ones for oblique incidence. We find this somewhat unsatisfactory and that it is avoidable if 
one consistently uses the travelling-frame convention. As demonstrated 
in \cite{remotecontrol} it is indeed possible to 
derive correct results if one carefully keeps track of the phase conventions. However both 
in regard to pedagogy and applications it 
would be desirable and simpler if a consistent convention were used and all 
reflections described similarly. If the travelling frame 
convention is used consistently the 
matrices for reverse propagation are of course n-transpose of the corresponding matrices 
for forward propagation  instead of being the transpose \cite{transpapp}.  
Though less familiar, the n-transpose is, however, an equally simple and elegant 
mathematical construct that satisfies the property $\overline{(AB)}=\bar B \bar A$.

The use of the  fixed-frame convention for reflection amplitudes has sometimes been 
justified by arguing that 
for normal reflections off an optically isotropic surface it yields a unit reflection 
matrix which avoids the asymmetry between the $s$ and $p$ wave reflection amplitudes. 
We point out that this is achieved at the expense of counter-intuitive behaviour of the 
amplitudes elsewhere. For example for reflection off ideal metallic mirrors at grazing incidence 
the fixed-frame convention gives a unit matrix suggesting no polarization 
change. We know however that under these conditions a right circularly polarized 
wave is reflected as a left circularly polarized wave and vice versa. Another problem 
with the use of the fixed-frame convention is that there is an asymmetry of 
conventions between the transmitted and the reflected waves. In a scattering problem 
there is no natural place for such an asymmetry. The neat correspondence between 
the theory of scattering of polarized waves and that of reflection and refraction is 
thus needlessly given up.

To sum up, in the examples discussed above we found cases 
(\cite{sosnowski}, \cite{engelsen} and \cite{silverman}) where the derived expressions satisfy 
the  constraints derived in this paper.
We found cases (\cite{lekner} and \cite{bassiri}) where they do so after accounting for a difference in 
phase convention. Finally we found a case (\cite{georgieva}) where the 
derived expressions donot satisfy the constraints  
and  we conclude that the derivation is incorrect.  
We wish to emphasize however that the 
satisfaction of the constraints is a necessary but not a sufficient condition for the 
correctness of the  derived reflection amplitudes. 
The constraints therefore provide only a partial test for the derived amplitudes. 
Finally we note that all the considerations in this paper relate to the linear 
regime of optics and do not include nonlinear phenomena.

\section{Acknowledgements}

I thank the referees for their comments which resulted in an enlarged, hopefully 
more convincing paper.

\clearpage

\begin{figure*}
\centerline{
\epsfxsize=0.7\textwidth
\epsfbox{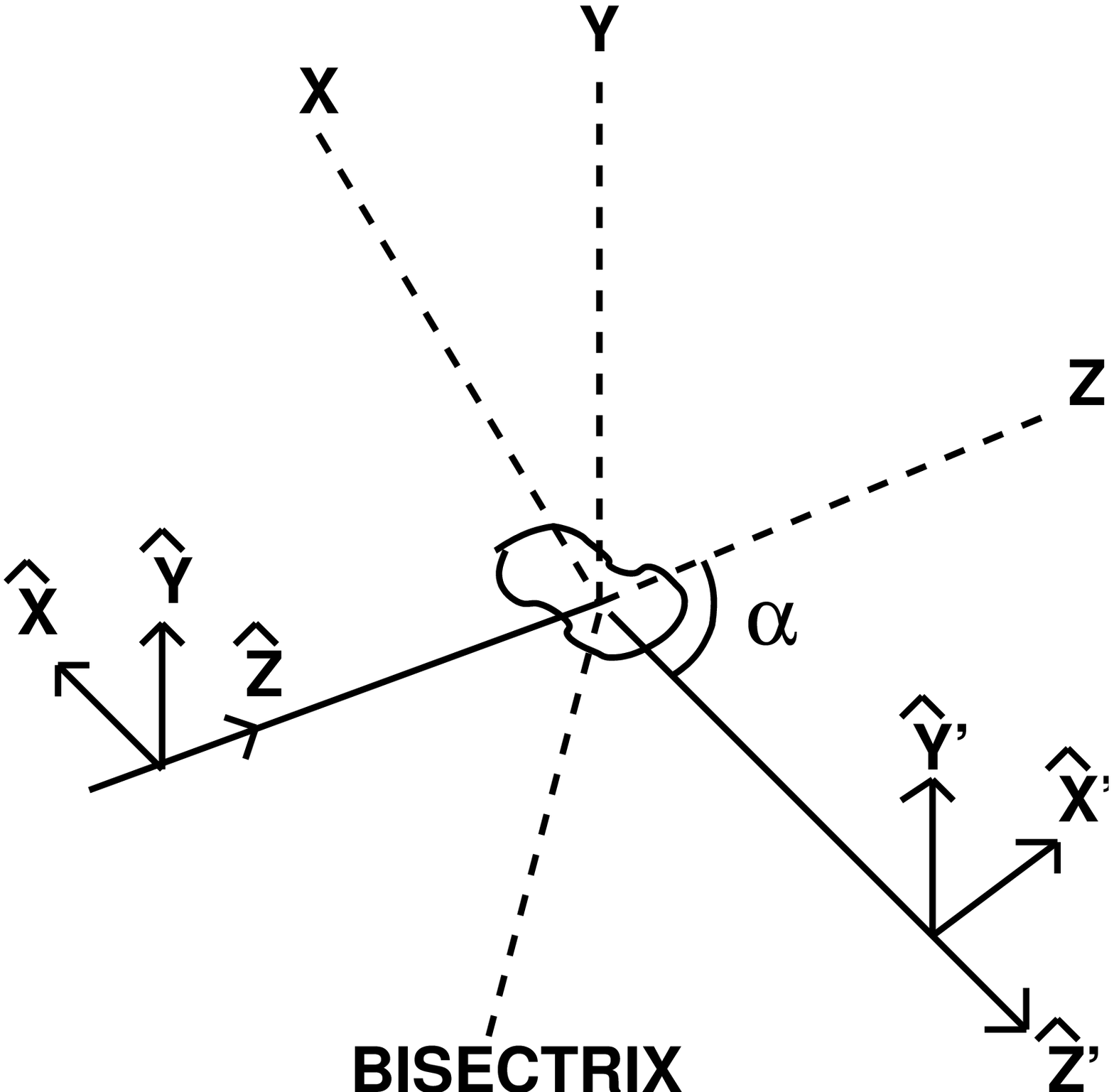}
}
\caption{The geometry of scattering of a plane wave with wave vector $\vec k$ along 
$\hat z$ to a wave with wave vector $\vec k'$ along $\hat z'$ where $\hat z$ and 
$\hat z'$ lie in the X-Z plane. The coordinate system ($\hat x'$,$\hat y'$,$\hat z'$) 
defining the polarization basis states in the scattered wave is obtained from the 
($\hat x$,$\hat y$,$\hat z$) system in the incident wave by a rotation about $\hat y$ 
through an angle $\alpha$ which is the scattering angle.}
\label{fig.1}
\end{figure*}

\begin{figure*}
\centerline{
\epsfxsize=0.7\textwidth
\epsfbox{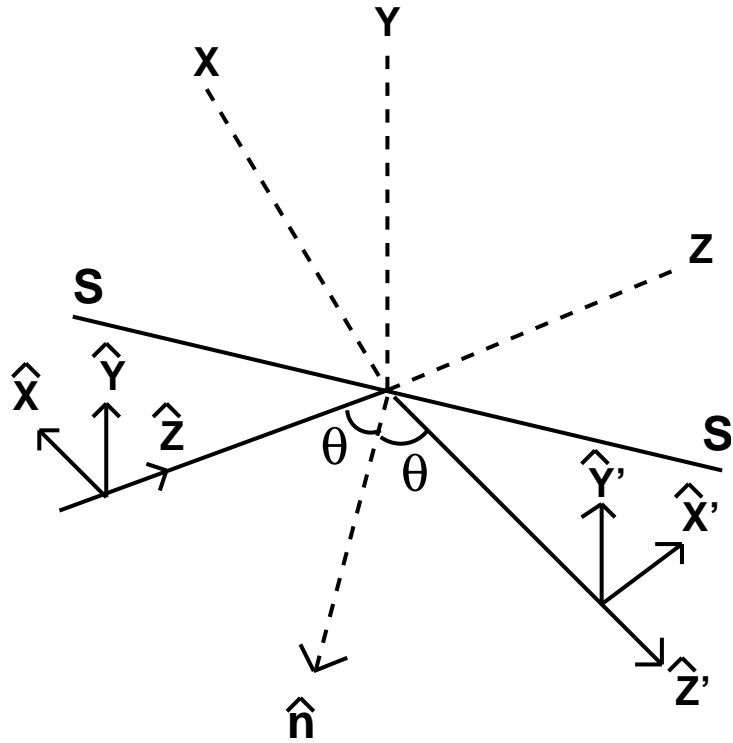}
}
\caption{The geometry of reflection of a plane wave propagating along $\hat z$ 
from a plane surface SS whose normal along $\hat n$ lies in the 
X-Z plane. The angle of incidence is $\theta$ and the relation between the coordinate 
systems ($\hat x$,$\hat y$,$\hat z$) and ($\hat x'$,$\hat y'$,$\hat z'$) is the 
same as in the scattering problem illustrated in Fig. 1.}
\label{fig.2}
\end{figure*}


\end{document}